\documentclass[aps,twocolumn]{revtex4}
\usepackage[utf8]{inputenc}
\usepackage{graphicx}
\usepackage{amsmath,color}
\usepackage{amsfonts}
\usepackage{nicefrac}
\usepackage{dcolumn}
\usepackage{bm}
\usepackage{ragged2e}
\usepackage{mathtools}
\usepackage{hyperref}
\hypersetup{
    colorlinks=true,
    linkcolor=blue,
    filecolor=mangeta,
    citecolor=blue,
    urlcolor=cyan
}
\begin{document}

\title{Tangential finite-size scaling of the Gaussian topological transition in the quantum spin-1 anisotropic chain}
\author{Luan M. Ver\'{i}ssimo, Maria S. S. Pereira, Marcelo L. Lyra}
\affiliation{Instituto de F\'{i}sica, Universidade Federal de Alagoas 57072-900 Macei\'{o} - AL, Brazil}

\begin{abstract}
Scaling aspects of Gaussian topological phase-transitions in quantum spin chains are investigated using the prototypical one-dimensional spin-1 XXZ Heisenberg model with uniaxial single-ion anisotropy $D$. This model presents a critical line separating the gaped Haldane and large-$D$ phases, with the relevant energy gap closing at the transition point. We show that a proper tangential finite-size scaling analysis is able to accurately locate the Gaussian critical line and to probe the continuously varying set of correlation length critical exponents. The specific features of the tangential scaling are highlighted in contrast with the standard scaling holding in the Ising-like transition between the gapless AF-Néel and gaped Haldane phases. Our results are compared with field-theoretic predictions and available high-accuracy  data for specific points along the Gaussian line.   
\end{abstract}

\maketitle
\section{Introduction}

Quantum spin models have been vastly investigated in several physical contexts. They are able to capture several interesting properties and outstanding collective behavior of strongly correlated electron systems. Interacting quantum spin models usually exhibit appealing physical phenomena, specially in one-dimension due to strong quantum fluctuation effects. The one-dimensional Heisenberg model describes localized spins interacting through an exchange coupling between neighbors. It can be viewed as the strong-interaction limit of a more complex Hubbard model for itinerant electrons\cite{Assa}. The isotropic antiferromagnetic Heisenberg model can exhibit distinct spin phases for integer or semi-integer spins. The spin-$\frac{1}{2}$ chain, for example, has a unique critical (massless) ground state, with gapless excitations and power-law decaying correlation functions. On the other hand,  the spin-1 chain has a disordered (massive) ground state, gaped excitations and exponential decay of correlations\cite{Assa,Hal}. The fundamental physical mechanism leading to such intriguing distinct behaviors was conjectured by Haldane in seminal works\cite{Haldane1,Haldane2}. Nowadays, the Haldane phase is understood as a symmetric protected topological phase of matter\cite{topo1,topo2}, characterized by a non-vanishing string order parameter\cite{VBS2} and the full breaking of a  hidden $ \mathbb{Z}_2 \times \mathbb{Z}_2$ symmetry\cite{Kennedy}.

Since Haldane's conjecture, quantum spin-1 chains, as well as several closely related models\cite{AffleckReview}, have been deeply studied. A rich scenario of quantum phases of matter has emerged with quantum phase transitions as a central issue\cite{QPT1,QPT2,QPT_Vojta}. An interesting model, which supports the Haldane's conjecture, was introduced by Affleck, Kennedy, Lieb, Tasaki (AKLT)\cite{AKLT,VBS} where the ground state was obtained exactly by mapping the spin-1 operator into the subspace of two spin-$\frac{1}{2}$ for each site. The ground-state spin configuration follows the construction of valence-bound states(VBS). The AKLT model features a Haldane gaped ground-state and can be interpreted as a particular case of a more general spin-1 model with bilinear and biquadratic interactions\cite{Bilinear2,Bilinear}. It features a phase diagram with ferromagnetic, dimerized, Haldane and critical phases. Another important class of quantum spin model emerges when one considers two kinds of anisotropies "Ising" and uniaxial single-ion, parameterized by $\lambda$ and $D$ respectively. The Heisenberg XXZ spin-1 model with single-ion anisotropy was extensively studied\cite{Botet,GS1,Schulz86,Golineli,PD,Onc=1,QMC1,Satoshi}. Its zero-temperature phase diagram is well understood nowadays, featuring an interesting type of topological quantum phase transition between two gaped phases (Haldane and large-$D$) separated by a gapless critical point that belongs to the Gaussian universality class. 

Indeed, the Gaussian class of topological quantum phase transitions occurs in several interesting interacting quantum spin model. However, the associated critical exponents are not universal. As such, the determination of critical Gaussian points and exponents is a challenging task. The twisted boundary conditions technique\cite{Kitazawa1}, for example, was implemented to determine the Gaussian critical point that separates Haldane and large-$D$ phases in a Heisenberg spin-1 XXZ chain with single-ion anisotropy. This technique explores the different values of eigenvalues of  space inversion and time-reversal operators in these phases\cite{PD}. The so-called multi-target density matrix renormalization group method\cite{MTDMRG} combined with finite-size scaling theory\cite{Onc=1} was also implemented in the same model, focusing in the precise determination of the Gaussian critical line. Another interesting aspect is the possibility to investigate Gaussian quantum phase transitions in a one-dimensional lattice model through a mapping transformation onto a Gaussian free two-dimensional field theory, with deep connections with conformal field theories\cite{Onc=1}.
Recently, a rhombic-type single-ion anisotropy spin-1 model was considered\cite{rhombicA,rhombicB}, where the phase-diagram features three Gaussian critical points that was investigated using numerical density renormalization group method\cite{dmrg1,dmrg2} and level spectroscopy method\cite{LS1,Kitazawa1,Kitazawa2}, including a discussion of  topological aspects of the Gaussian transitions. 

From the quantum information perspective, several kinds of measurements of entanglement was proposed to localize critical points where the quantum phase transition occurs (see, e.g \cite{EntMB}). The fidelity approach\cite{Fidelity1,Fidelity2} was implemented to localize the critical points in anisotropic spin-1 model, where the fidelity susceptibility fails to detect the Gaussian transition in $\left( D - \lambda \right)$ plane for $\lambda = 0.5$. The scaling behavior of von Neumann entropy has been used to determine accurately the central charge of the associated conformal field-theory and the phase boundaries of the Gaussian topological quantum phase transitions in several models including an extended Bose-Hubbard\cite{c1,c2,c3}, dimmerized spin-1 XXZ with uniaxial single-ion anisotropy\cite{satoshi2}, as well as topological aspects of spin-2 quantum chains\cite{roman1,roman2}. An improved error protocol method for DMRG was introduced\cite{Gaussian1}, where the critical Gaussian points was obtained with high accuracy exploring the von Neumann entropy behavior for large system sizes. Recently, a finite-size scaling analysis of the energy gaps in the vicinity of Gaussian transitions was introduced\cite{Luan1,Luan2} in prototype models of heterotrimetallic compounds described as branched chains with alternate $S=1/2$ and larger spins.  It was shown that the closing of the gap on both phases meeting at the Gaussian critical point requires a proper tangential scaling form. Such adapted tangential finite-size scaling analysis was shown to provide accurate estimates of the Gaussian critical point, as well as of the associated correlation length critical exponent.   

In this work, we aim to detail the tangential finite-size scaling technique and to demonstrate its capability and accuracy. To reach this goal, we will apply this technique to investigate the quantum critical behavior of the spin-1 Heisenberg XXZ chain with uniaxial single-ion anisotropy. This model is known to display a Gaussian critical line with continuously varying critical exponents\cite{PD,Schulz86}. Using a proper tangential scaling form of the scaled gap in the vicinity of the Gaussian quantum critical points, we will estimate their location along the full transition line. Collapse of data obtained from density-matrix renormalization group calculations on finite chains into the tangential scaling form will be used to provide accurate values of the correlation length critical exponent along the Gaussian line. The estimated correlation length critical exponents will be compared to analytical predictions based on the mapping transformation in an effective continuum O(2) nonlinear field-theoretic  nonlinear $\sigma$ model\cite{Onc=1,Low}, as well as with previous high-precision numerical estimates\cite{Gaussian1}.

\section{Model and the $O(2)$ nonlinear $\sigma$ model mapping} 

The XXZ spin-1 chain with uniaxial single-ion anisotropy is described by the following Hamiltonian : 

\begin{eqnarray} \label{eq1}
\mathcal{H} &=& \sum_{i=1}^{N} J \left[\left(S_{i}^{x}S_{i+1}^x + S_{i}^{y}S_{i+1}^{y} +\lambda S_{i}^{z}S_{i+1}^z\right)\right] \nonumber \\  &+& D\sum_{i=1}^{N}(S_{i}^{z})^2
\end{eqnarray}

\noindent where $S_{i}^{\alpha}$ denotes the component $\alpha = x,y,z$ of the spin-1 operator. The parameters $\lambda$ and $D$ represent the "Ising-like" and uniaxial single-ion anisotropies respectively. The isotropic antiferromagnetic exchange parameter $J$ here will be fixed as the energy scale ($J = 1.0$). The ground-state phase diagram of the above model features several quantum phases\cite{PD} such as Haldane, Large-$D$, ferromagnetic, AF-Néel phases, besides two critical $XY$ phases distinguished by their low-lying spin excitations\cite{Schulz86}. Several kinds of quantum phase transitions takes place. Here we feature three of those: Gaussian topological transition between the gapful Haldane and large-$D$ phases that have opposite parity symmetry, Kosterlitz-Thouless transitions between  gapless XY and gapful Haldane or large-$D$ phases and Ising transition between Haldane and Néel phases. Here, we will focus in absolute positive $\lambda >0$ and $D>0$) parameters space of the phase diagram where the Haldane, Large-$D$ and AF-Néel phases appear. 
%As already mentioned, the determination of phase boundaries of Gaussian transitions was already studied from several ways and above we bring some quantum field theory ideas we can use.

From the quantum field theory point of view, the lattice spin-1 model described by  Eq.\eqref{eq1} along the Gaussian transition critical line can be mapped onto a free Gaussian model in Euclidean space  by using the path-integral representation of the partition function and  spin coherent states \cite{Assa}. The procedure described in Ref. \cite{Onc=1,Low} leads to an effective continuum $O(2)$ nonlinear $\sigma$ model (NL$\sigma$M) : 

\begin{equation}
    \mathcal{L}_{O(2)} = \frac{1}{2}\left[ \frac{1}{v} (\partial_{\tau}\Theta)^2 + v(\partial_{x} \Theta)^2  \right]
    \label{eq2}
\end{equation}
and
\begin{eqnarray}
\label{eqg}
g = \frac{1}{s}\sqrt{2(1+D+\lambda)} \\ 
v = s \sqrt{2(1+D+\lambda)}
\end{eqnarray}

\noindent with $\Theta$ the bosonic field compactified in a circle of radius $\frac{1}{\sqrt{g}}$ and $v$ is the spin-wave velocity of the theory. This procedure establishes a direct connection of the lattice model parameters with $g$ and $v$ of the continuum model which allows us to compare numerical calculations for this lattice model with the field theoretical predictions. It is described by the conformal field theory with central charge $c=1$\cite{Low} and a nonuniversal radius compactification parameter $K = \pi/g$. The cases $K=1/2$ and $K=1$ correspond to special self-dual and free Dirac points\cite{Onc=1,satoshi2}. The Kosterlitz-Thouless quantum phase transition has $K=2$ as demonstrated, for example, in the Tomonaga-Luttinger regime of anisotropic spin chains\cite{Rene} after a phenomenological bosonization transformation\cite{Guiamarchi}.

\section{Finite size-scaling: Ising and Gaussian points}

The scaling behavior of excitation energy gaps is usually explored in the determination of phase boundaries and critical points in several quantum spin models\cite{ODM}. However, some challenging aspects are involved in non-conventional classes of quantum phase transitions, such as Kosterlitz-Thouless and Gaussian critical points. The scaling form of excitation gaps depends of the own nature of the specific quantum phase transitions. For example, the Ising class features a crossing behavior of the scaled excitation gap curves in the vicinity of the critical point. On the other hand,  the scaled excitation gap curves behaves differently in Gaussian critical points, presenting a tangential scaling in the vicinity of the transition\cite{Luan1,Luan2}. Exploring these ideas, we develop a simple method to determine the critical points and correlation length critical exponents directly from the scaled excitation gap data for the Gaussian class of quantum phase transitions. Data were obtained by employing the tensor network\cite{Orus,Ulrich} formulation of  numerical density matrix renormalization group method(DMRG)\cite{dmrg1,dmrg2}. All data we are going to report were obtained using the open-source code from the Algorithms and Libraries for Physics Simulations(ALPS) project \cite{ALPS}. We considered periodic-boundary conditions and several system sizes up to $N = 100$. The bond-dimension $\chi$ inheent of DMRG calculations was chosen seeking truncation errors of the order of $10^-9$.

In general, the spin gap $E_s$ is defined by: 

\begin{equation} \label{eq3}
    E_s = E_0(1,N) - E_0(0,N)
\end{equation}

\noindent where $E_0(M,N)$ is the ground-state energy  with $N$ spins in the total magnetization sector $M = \sum_i^N S^z_{i} $. Since Haldane and large-$D$ are gaped phases, we expect the spin gap to remain finite in both phases. At the Gaussian transition point, such energy gap decreases as $1/N$ with increasing system sizes. 

The AF-Néel to Haldane phase transitions belongs to the Ising class of phase transitions. The proper quantity to explore such critical point is the neutral gap $E_n$ given by: 

\begin{equation}
    E_n = E_1(0,N) - E_0(0,N)
\end{equation}

\noindent corresponding to the difference between the first excited and ground state energies in the total magnetization  sector $M=0$. Here the energy gap differs from the spin gap in the quantum number $M$, calculating the energies difference at the same magnetization sector for two low-lying eigenvalues. While in the Haldane phase this energy gap remains finite in the thermodynamic limit, it vanishes as $1/N$ at the Ising critical point and faster within the AF-Néel phase.

Exploring the vanishing behavior of the energy gaps at the critical points for several systems sizes, we expect the scaled  gaps $\Delta_s = NE_s$  or $\Delta_n = NE_n$ curves for diverse systems sizes to meet together only at the critical point. Exploring this feature, we built the phase diagram of the model, as reported in Fig.\ref{figA}. Details of the scaling procedure will be given below. Our results give reliable estimates of some representative features of the phase diagram that have been previously raised using other techniques\cite{PD}. For example, the extrapolation to the tricritical point where the Gaussian and Ising lines meet is $(D \simeq \lambda \sim 3.0)$. Further, we could capture with good agreement the critical point $D_c = 0.3493$ at the axis $\lambda = 0$ corresponding to the interface of Haldane, Large-$D$ and $XY$ phases. This point represents a Kosterlitz-Thouless transition between a critical and a gaped phase. We also got the critical point $\lambda_c = 1.1862 $ at the axis $D = 0.0$ for the Ising curve separating Haldane and AF-Néel phases.

\begin{figure}[ht!]
    \centering
    \includegraphics[width=160pt, clip]{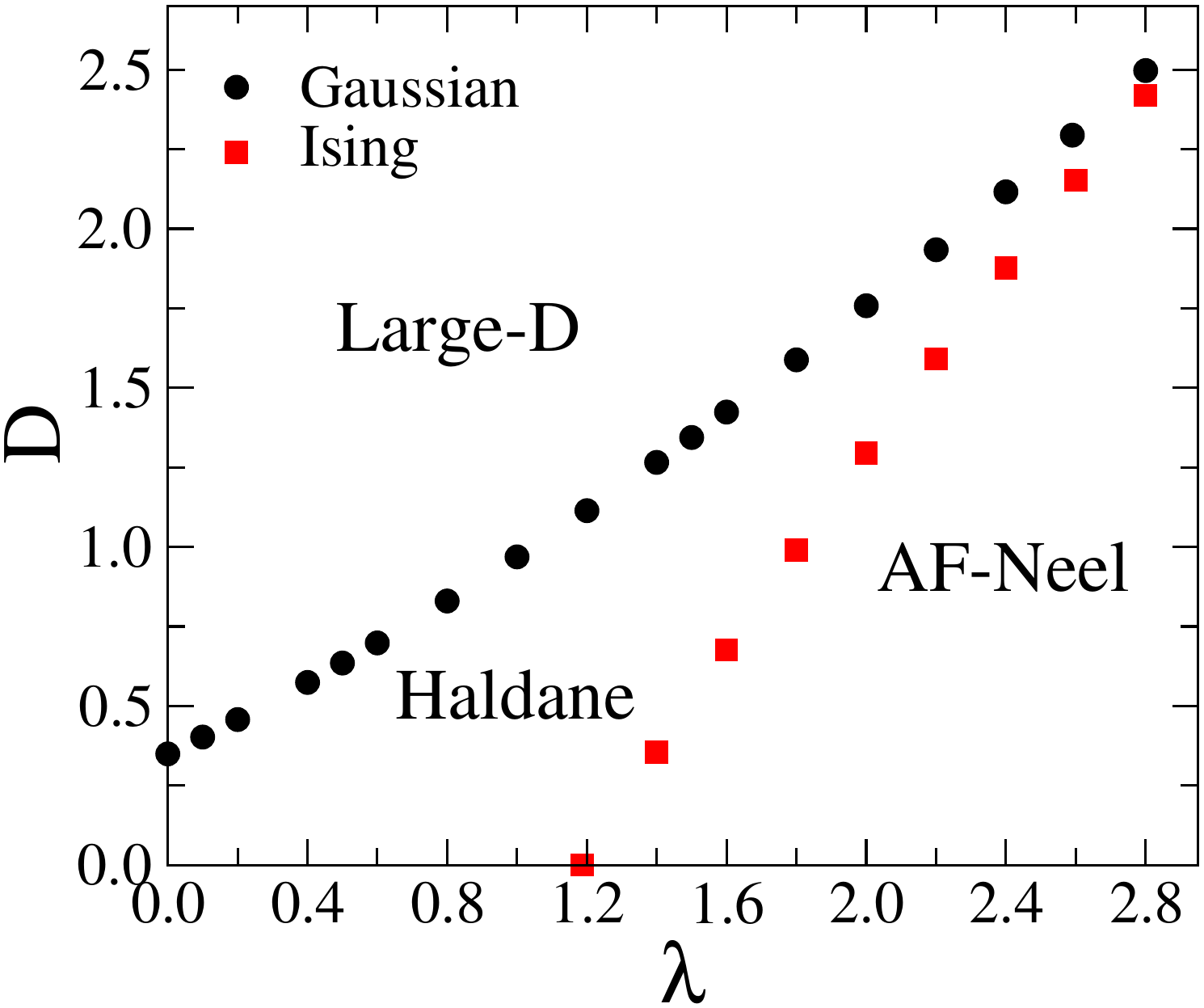}
    \caption{Ground state phase diagram of the Hamiltonian \eqref{eq1} for  positive $\lambda$ and $D$. Black circles denote the Gaussian transition critical line between Haldane and large-$D$. Red squares represent the Ising critical line separating Haldane and AF-Néel phases. }
    \label{figA}
\end{figure}

\subsection{Ising transition: standard finite-size scaling}

\begin{figure}
    \includegraphics[width=160pt, clip]{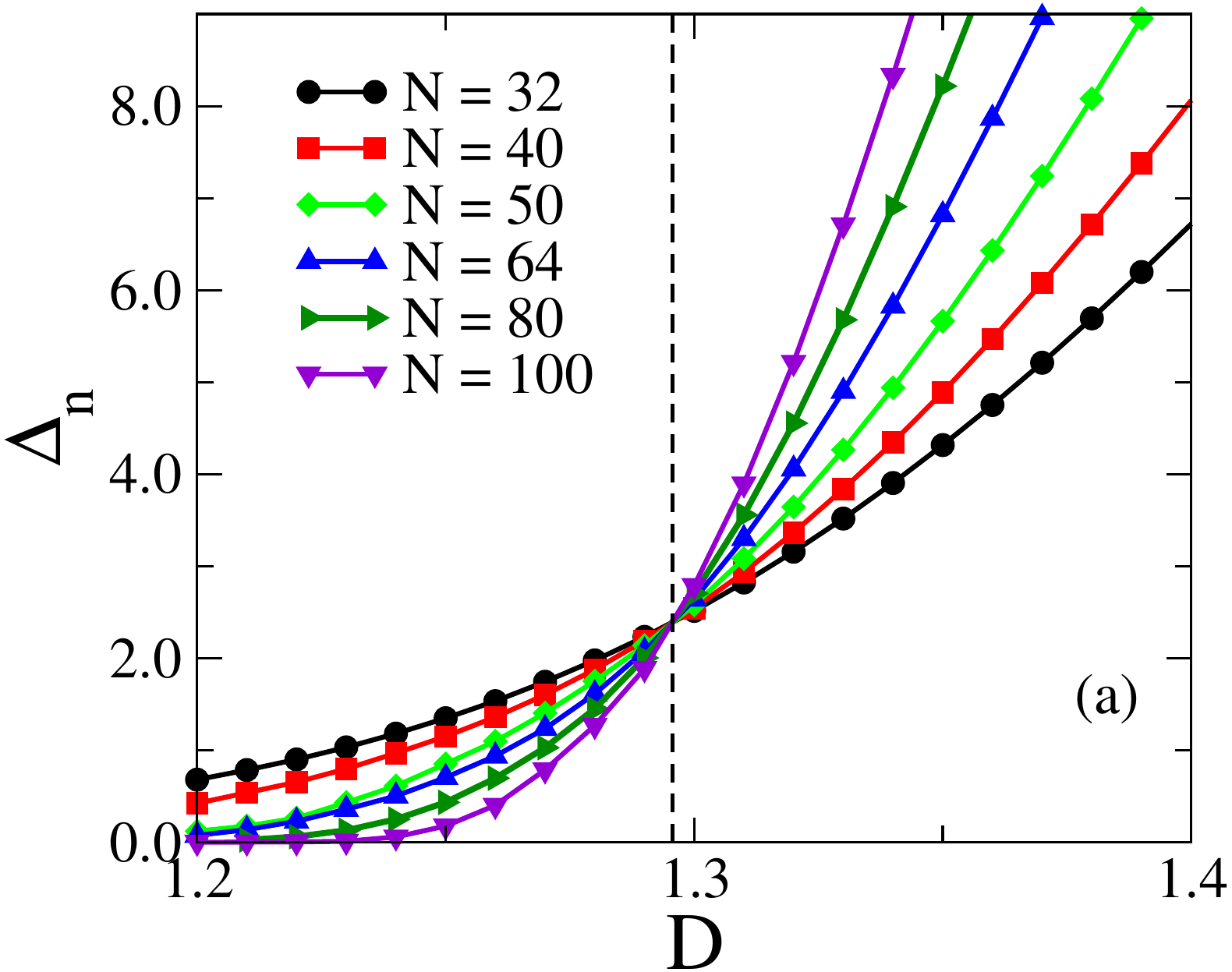}
    \includegraphics[width=160pt, clip]{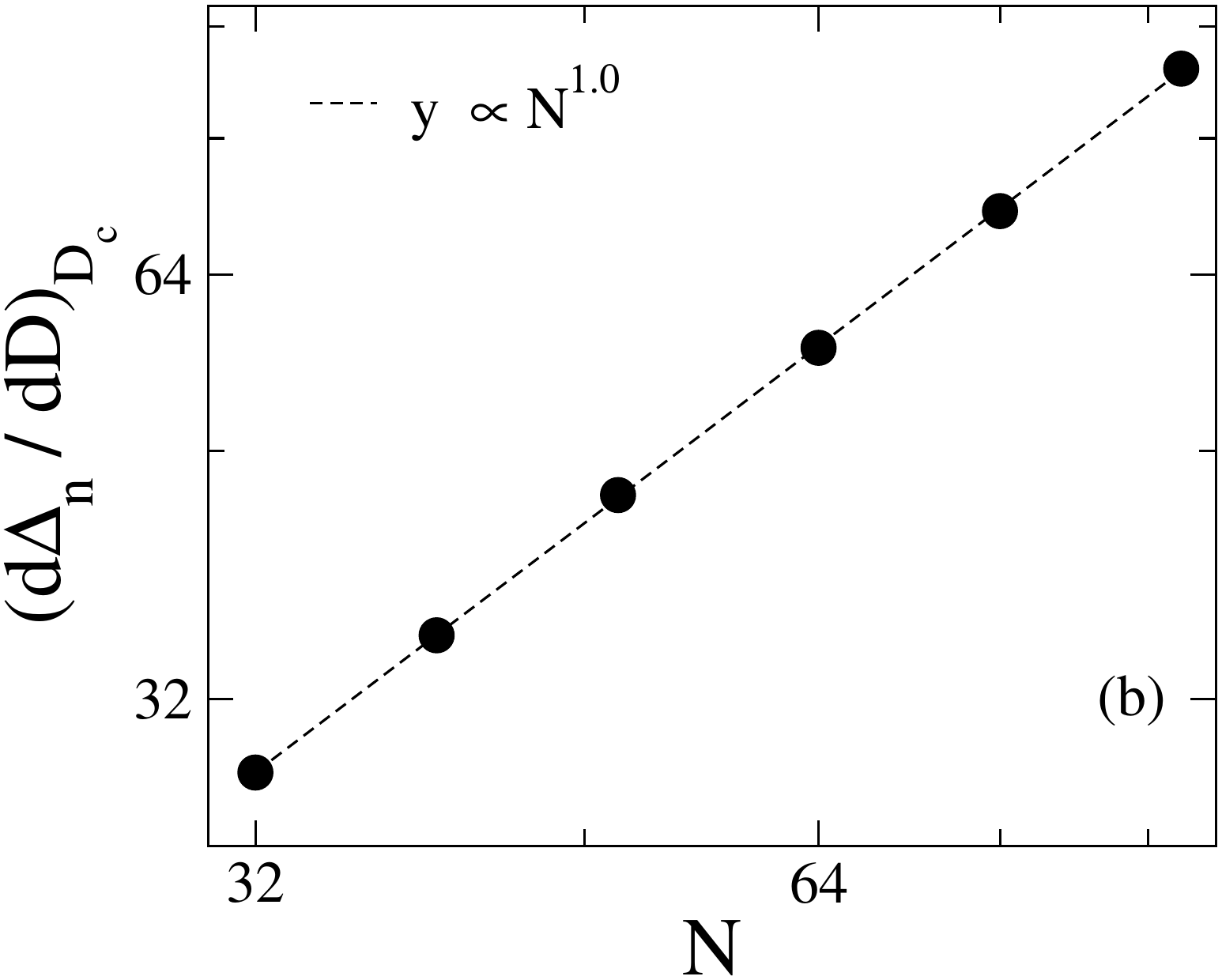}
    \includegraphics[width=160pt, clip]{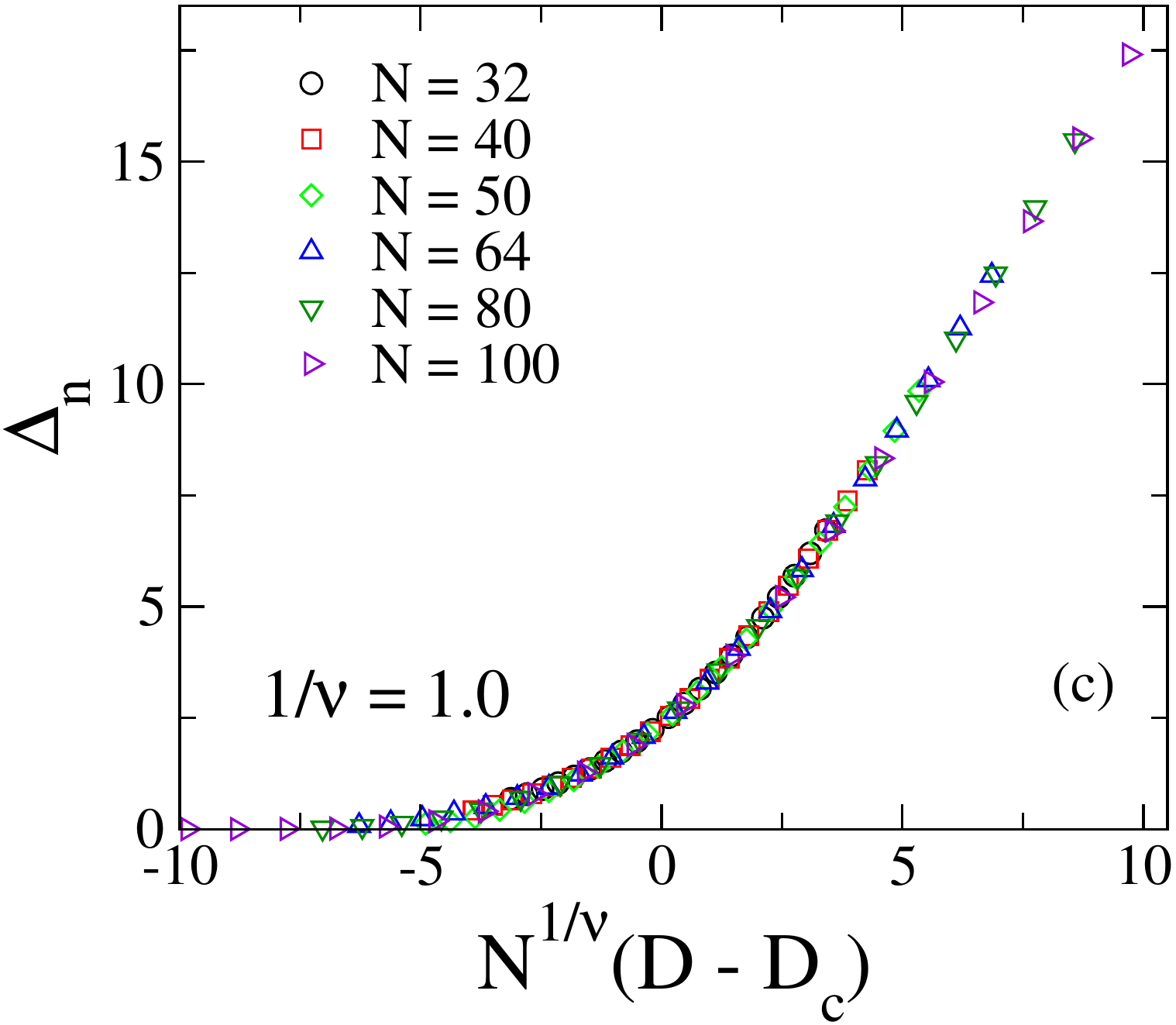}
    \caption{Standard finite-size scaling in the vicinity of  the Ising-like transition. In (a) we report the crossing behavior of the scaled neutral gap $\Delta_n$ in the  vicinity of the critical point $D_c$. In (b) we plot $d\Delta_n/dD$ at the critical point as a function of the chain size $N$ which is consistent with the correlation length critical exponent  $1/\nu=1$. Figure (c) shows the universal scale invariant behavior of the scaled neutral gap against the proper scaling variable, endorsing the accuracy of our estimates of critical point and critical exponent and the validity of the single parameter scaling hypothesis. }
    \label{figB}
\end{figure}

We start the  description of the finite-size scaling technique by recalling its main features that hold in the vicinity of the usual Ising phase transition between Haldane and AF-Néel phases. The correlation length $\xi$ is the only relevant length scale in the thermodynamic limit that diverges at the critical point. For a fixed value of $\lambda$, the correlation length scales as $\xi\propto |D-D_c|^{-\nu}$   with the well known critical exponent $\nu = 1.0$. For finite chains, the single parameter scaling hypothesis implies that the relevant physical quantities shall depend only of the ratio $\xi /N$, at least in the vicinity of the critical point where $\xi>>1$.  According to this reasoning, data of the scaled neutral gap $\Delta_n$ obtained for distinct values of chain sizes and anisotropies, shall obey the universal scaling form: 

\begin{equation} \label{eq7}
\Delta_n = f \left( \frac{\xi}{N} \right)=g[(D-D_c)N^{1/\nu}]
\end{equation}

\noindent with $g(x\rightarrow - \infty)\rightarrow 0$. $g(x\rightarrow +\infty )\propto x^{\nu}$ warranties that the neutral energy gap becomes size independent in the thermodynamic limit within the Haldane phase.  To account for the size independence of the scaled neutral gap at the transition, one might have $g(0)$ being a constant. At the critical point, the first derivative of (\ref{eq7}) scales as $d\Delta_n/dD \propto N^{1/\nu}$. The main steps of the above finite-size scaling analysis are summarized in Fig.($\ref{figD}$). These results were obtained for $\lambda = 2.0$ and sweeping the values of $D$ around the critical point. First, we explore the crossing behavior of the scaled gap $\Delta_n$ curves for several systems sizes which allows us identify the critical point $D_c=1.2956()$ in Fig. (\ref{figB})(a). After that, we compute the first derivatives of $\Delta_n$ at $D = D_c$ for each chain size. When plotted as a function of $N$  in log-log scale, as shown in Fig. ($\ref{figB}$)(b), the slope gives the correlation length critical exponent $1/\nu = 1.0$. Further, we evaluate the accuracy of  the above estimates of the critical point and correlation length exponent by plotting the scaled neutral gap $\Delta_n$ versus the proper scaling variable using the scaling form for the Ising universality class given by Eq.\ref{eq7}.  The collapse of all data from  distinct chain sizes into a single curve corroborates the estimated critical parameter and the validity of the single parameter scaling hypothesis. The above procedure was performed along the entire Ising critical line. As expected, the correlation length exponent $\nu=1$ was found to be universal.

\subsection{Gaussian transition: tangential finite-size scaling}

The Néel-Haldane phase transition discussed above involves phases with null (AF-Néel) and diverging (Haldane) scaled neutral gaps as $N \rightarrow \infty$. In contrast, the proper scaled spin gap diverges on both Haldane and large-$D$ phases. Scale invariance still holds at the Gaussian transition point between these phases. Therefore, a new scaling analysis shall be developed for this specific phase transition for which curves of the scaled gap from distinct system sizes do not cross but rather touch each other tangentially at the critical point.

\begin{figure}
     \centering
    \includegraphics[width=160pt, clip]{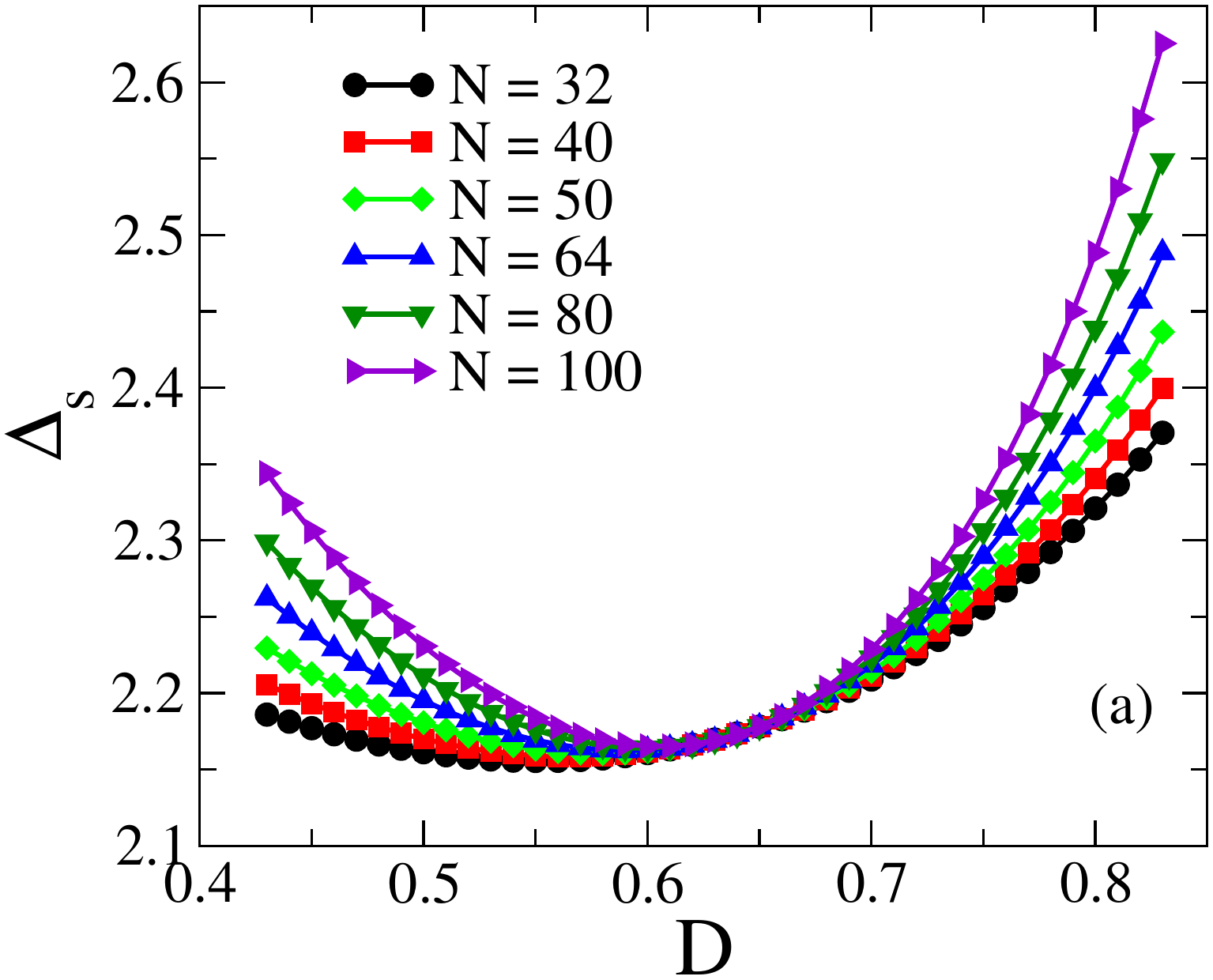}
    \includegraphics[width=160pt, clip]{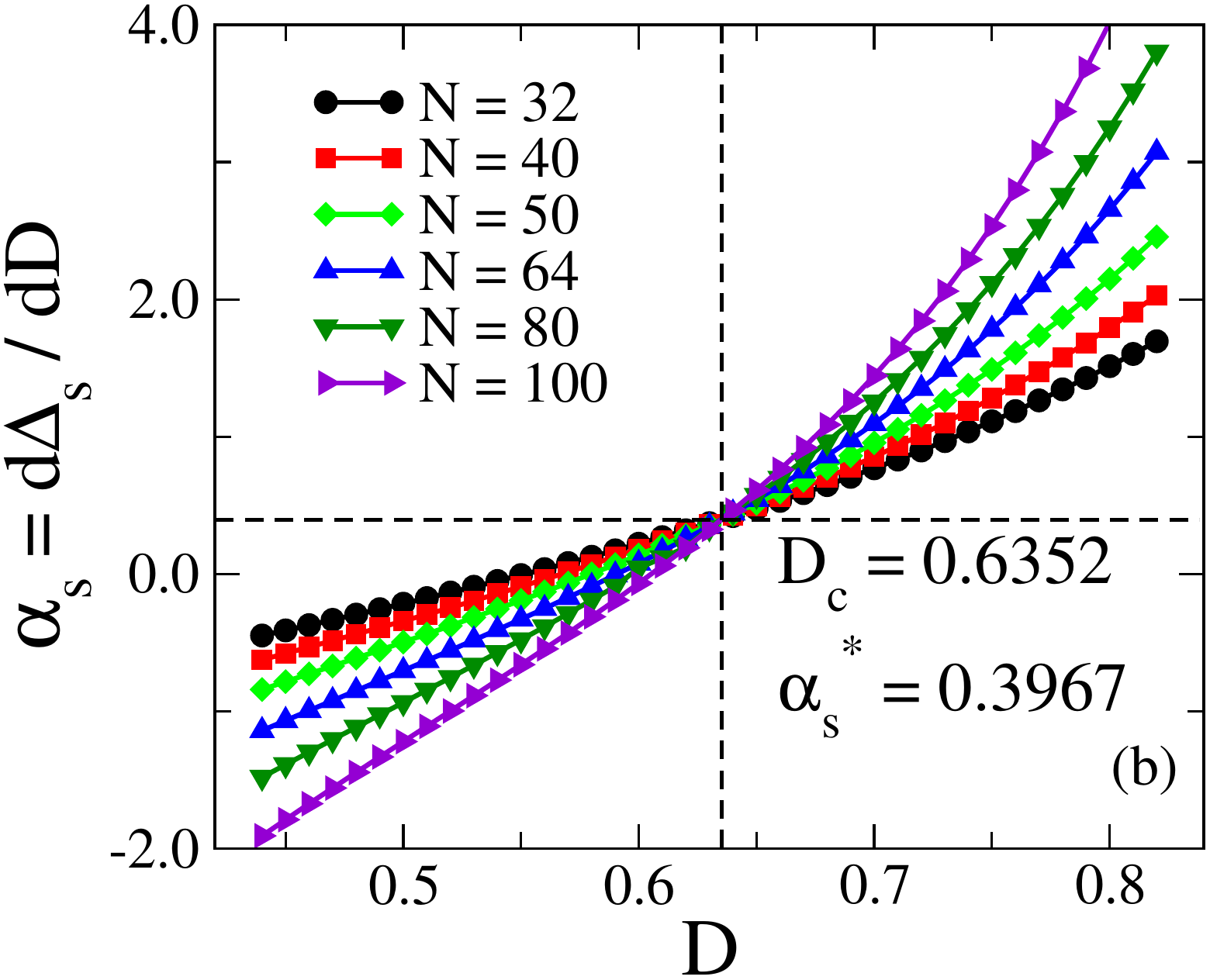}
    \includegraphics[width=160pt, clip]{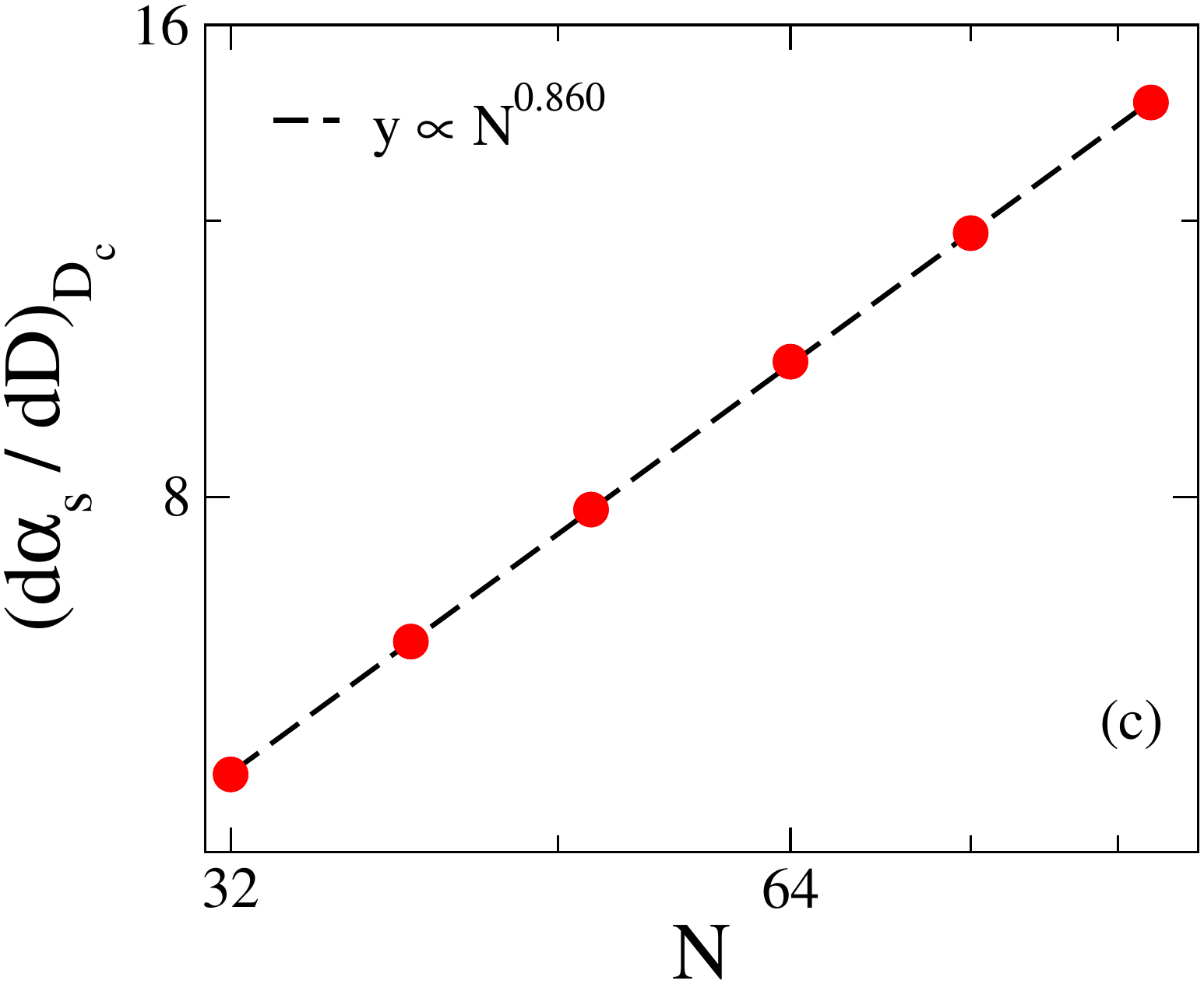}
    \includegraphics[width=160pt, clip]{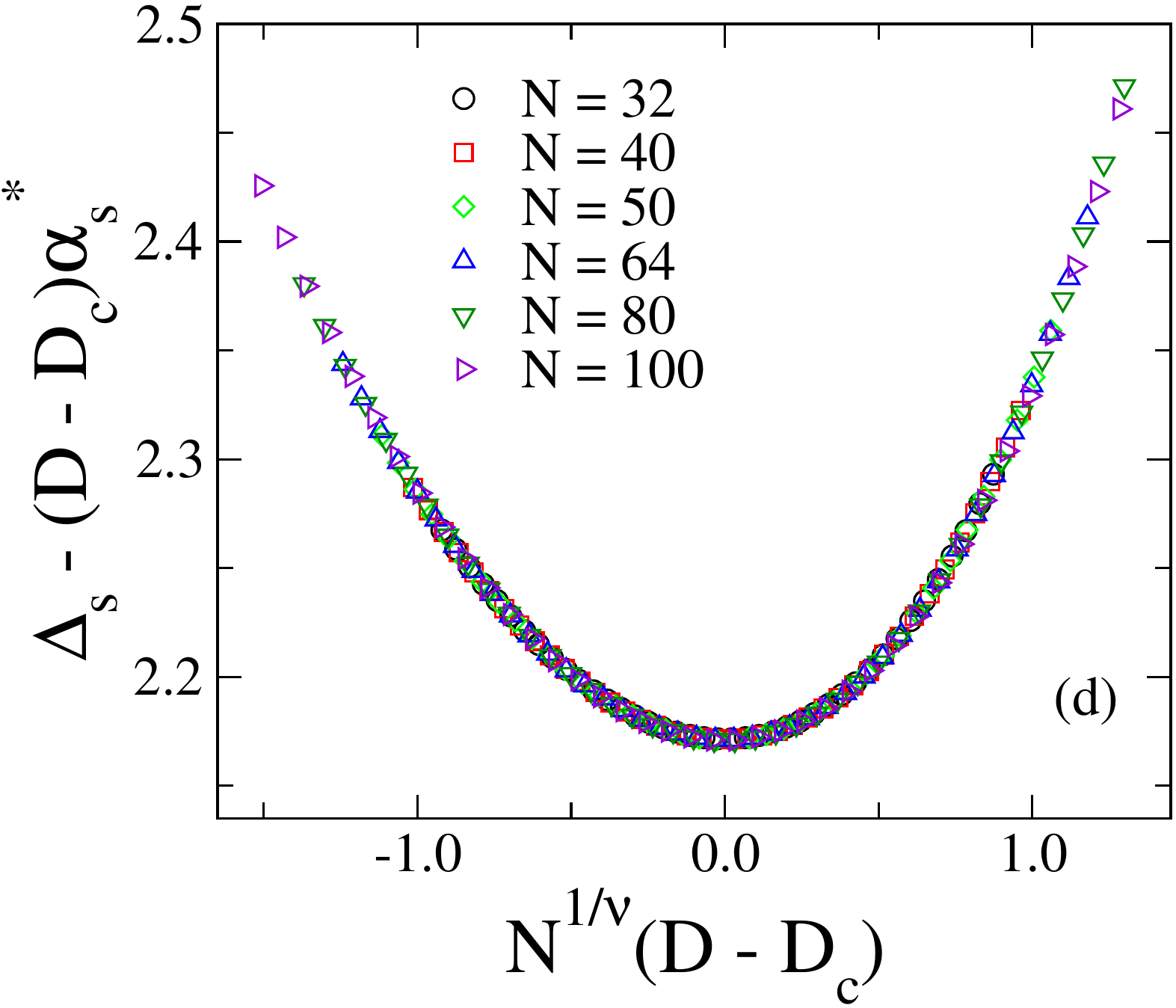}
    \caption{Figure (a) shows the scaled spin gap $\Delta_s$ for $\lambda = 0.5$ considering several system sizes around the Gaussian critical critical point. The tangential behavior at the critical point is evidenced. Figure (b) features the first derivative of scaled gap ${\alpha}_s$. The crossing point identifies precisely the Gaussian critical point $D_c=0.6352(2)$ and $\alpha_s^*=d\Delta_s/dD|_c= 0.3967(6)$. In (c) the power-law behavior of the second derivative of the scaled gap at critical point  allows to extract the correlation-length critical exponent $1/\nu = 0.430(4)$. In Figure (d) we provide the data collapse of scaled spin gap curves using the estimated critical parameters onto the tangential finite-size scaling form.}
    \label{figC}
\end{figure}

Let us describe in detail the tangential finite-size scaling procedure that is appropriated  to characterize the Gaussian critical point. Using the fact that the energy gap $E_s$ closes just at the critical point, the scale invariant behavior of the system at the quantum Gaussian critical point emerges naturally. We will illustrate the method for two points along the Gaussian line, namely, $\lambda = 0.5$ and $\lambda = 2.0$. The first step of the method is to compute the scaled spin gap curves for several systems sizes in the vicinity of the Gaussian transition point between Haldane and large-$D$ phases at a given $\lambda$ by sweeping the anisotropy $D$ values. We can see in Figures \ref{figC}(a)  and \ref{figD}(a) the typical behavior of the scaled spin gap. In each case, we used data from several system sizes up to chains with $N=100$ spins. The scaled spin gap $\Delta_s$ indeed exhibits a quite different size-dependence as compared to the scaled neutral gap $\Delta_n$ for the Ising transition. Instead of crossing at a single point, curves from distinct system sizes touch tangentially at the critical Gaussian point, as previously inferred. Therefore, not just the scaled spin gap but also its derivative ${\alpha}_s = d\Delta_s/dD$ become scale invariant at the Gaussian transition.

The second step is to precisely locate the Gaussian critical point computing numerically the first derivative of scaled spin gap ${\alpha}_s$  as a function of the anisotropy parameter $D$, as shown in Figures \ref{figC}(b) and \ref{figD}(b). These are the quantities that cross at the Gaussian critical point because the scaled spin gap and its derivative are both scale invariant but the curvature increases with the chain size. From the crossing at the critical points, we estimate $D_c = 0.6352(2)$ and $D_c = 1.7582(2)$ for $\lambda = 0.5$ and $\lambda = 2.0$ respectively.

To estimate the correlation length critical exponent, we need to extract the scale invariance of the first-derivative of the scaled spin gap at the critical point from the single parameter finite-size  scaling hypothesis. To account this feature, the finite-size scaling form of the scaled spin gap is written as

%valuate the validity of the method, we can exploit scaling properties of scaled spin gap at the critical point using an extended scaling form of scaled spin gap beyond the usual scaling hypothesis that consider the independence of $\alpha_s$ at the critical point for several systems sizes. In that way, considering the finite-size scaling effects we propose an proper form of scaled spin gap in vicinity of Gaussian critical point should hold the following form

\begin{eqnarray}
\Delta_s &=& (D-D_c){\alpha}_s^* + f \left( \frac{\xi}{N} \right) \nonumber \\
&=& (D-D_c){\alpha}_s^* + g[(D-D_c)N^{1/\nu}]
 \label{eq8}
\end{eqnarray}

\vspace{0.25cm}
\noindent where $g(x)$ is quadratic at $x=0$, diverging for $ x \rightarrow \pm\infty$. Here ${\alpha}_s^*=d\Delta_s/dD|_c$. This scaling form ensures that the first derivative of the scaled spin gap at the critical point is scale invariant. The second derivative $d^2\Delta_s/dD^2 = d{\alpha}_s^*/dD = d^2 g/dD^2$ computed at the Gaussian critical point becomes proportional to $N^{2/\nu}$. The third step of the method, therefore, corresponds to plot these values as  function of the system size and to extract the correlation-length critical exponent, as shown in Figures $\ref{figC}$(c) and $\ref{figD}$(c). Notice that the correlation-length critical exponent is non-universal, as expected along the Gaussian critical line\cite{PD,Onc=1}.

To complete the tangential finite-size scaling analysis, we perform a data collapse to evaluate the accuracy of the single-parameter scaling hypothesis Eq.\ref{eq8}. The proper quantity to be rescaled is $\Delta_s - \left(D - D_c\right) \alpha_s^*$. Their values, when plotted against $x=N^{\frac{1}{\nu}}\left(D - D_c\right)$, shall collapse  onto an single universal curve $g(x)$ independent of the chain size. Such data collapse process is shown in Fig.\ref{figC}(d) for $\lambda = 0.5$ and Fig.\ref{figD}(d) for $\lambda = 2.0$ using the critical parameters estimated from the previous steps. The quadratic form of the scaling function $g(x)$ at $x=0$ is a signature of the tangential finite-size scaling.

\section{Discussion}

\begin{figure}
    \centering
    \includegraphics[width=160pt, clip]{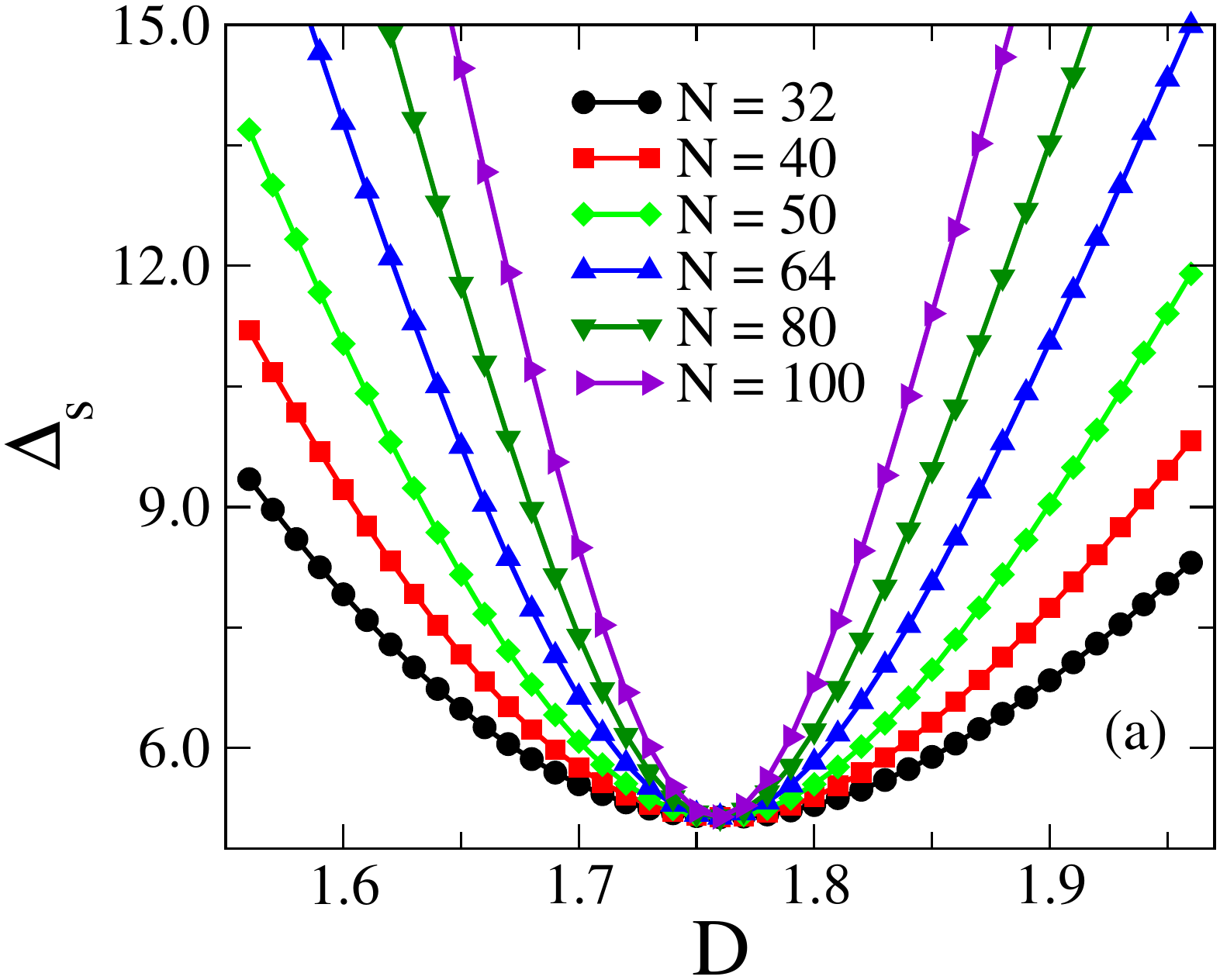}
    \includegraphics[width=160pt, clip]{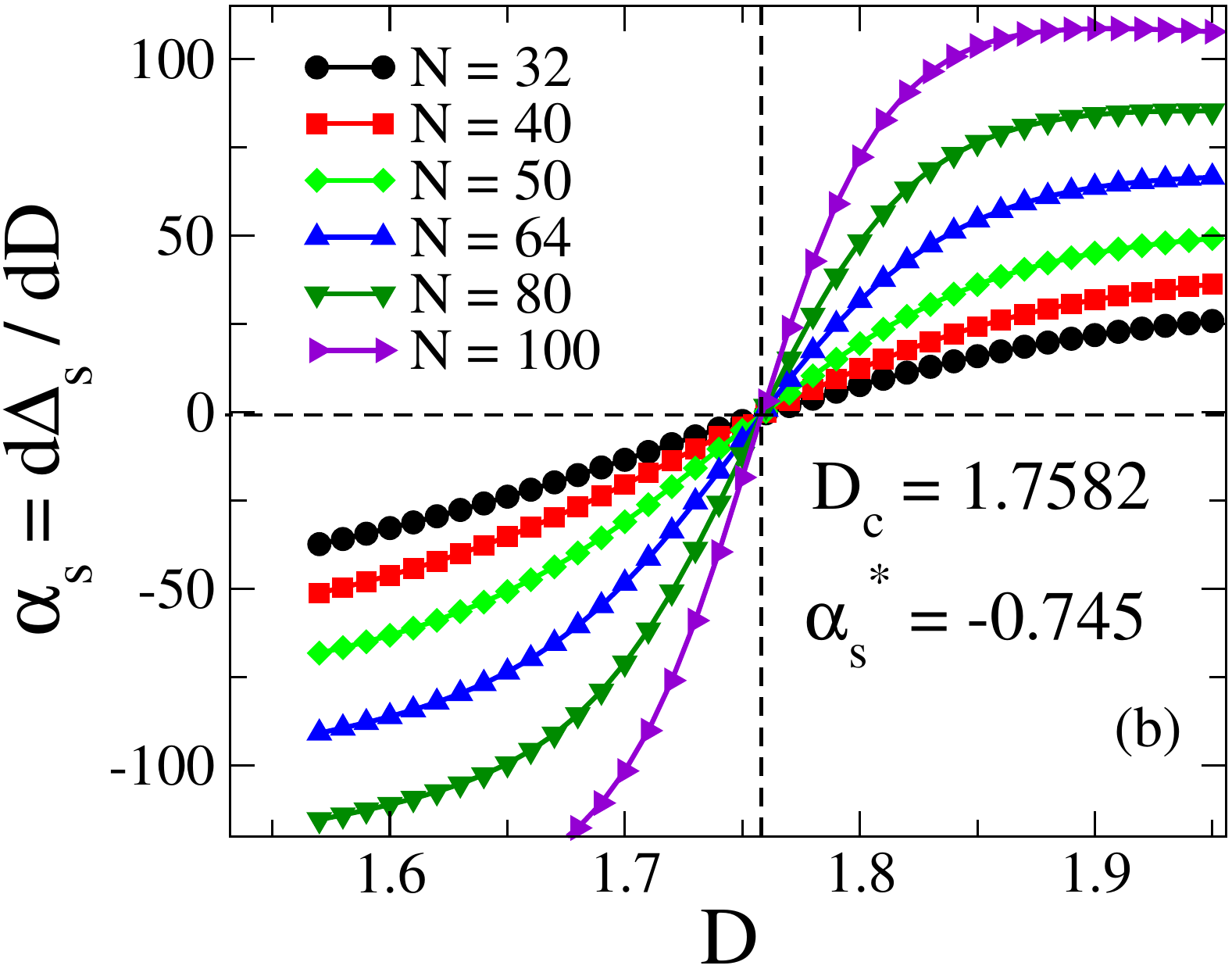}
    \includegraphics[width=160pt, clip]{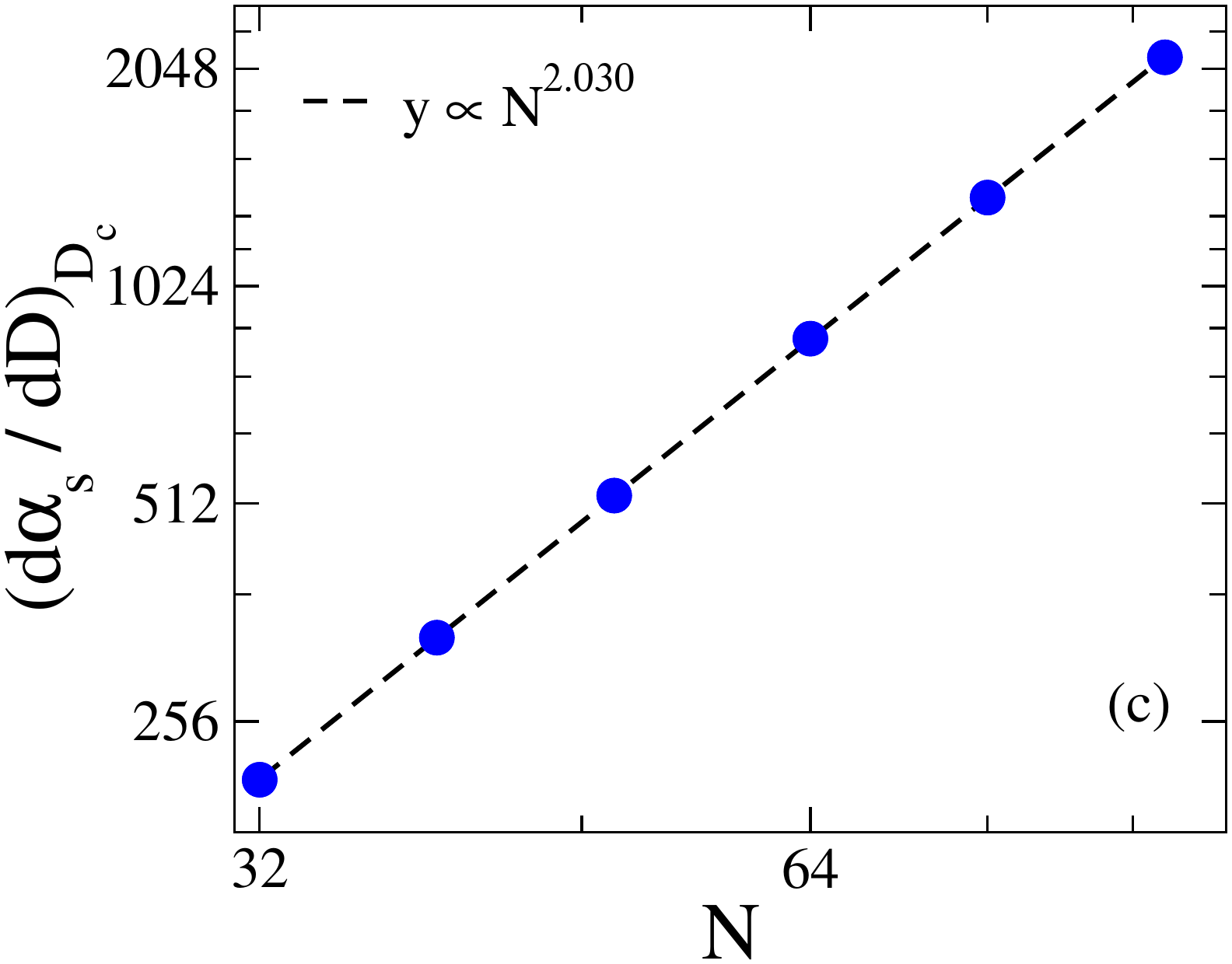}
    \includegraphics[width=160pt, clip]{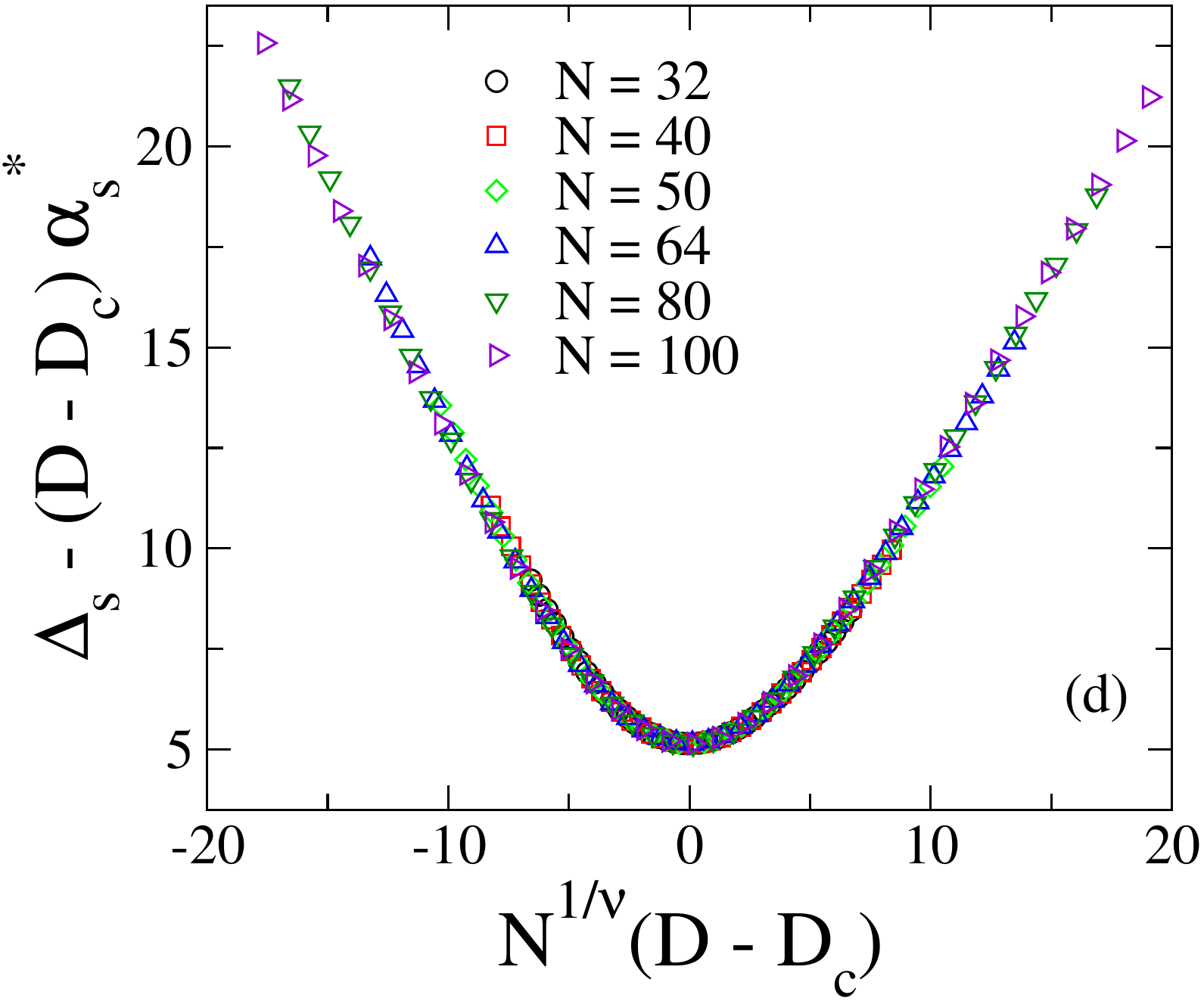}
    \caption{Figure (a) shows the scaled spin gap $\Delta_s$ for $\lambda = 2.0$ considering several system sizes around the Gaussian critical critical point. Figure (b) features the first derivative of scaled gap ${\alpha}_s$. The crossing point identifies the Gaussian critical point $D_c=1.7582(2)$ and $\alpha_s^*=d\Delta_s/dD|_c=-0.745(9) $. In (c) the power-law behavior of the second derivative of the scaled gap at critical point  allows to extract the correlation-length critical exponent $1/\nu = 1.015(5)$. In Figure (d) we provide the data collapse of scaled gap curves using the estimated critical parameters onto the tangential finite-size scaling form.
    }
    \label{figD}
\end{figure}

\begin{figure}[ht!]
    \centering 
    \includegraphics[width=160pt, clip]{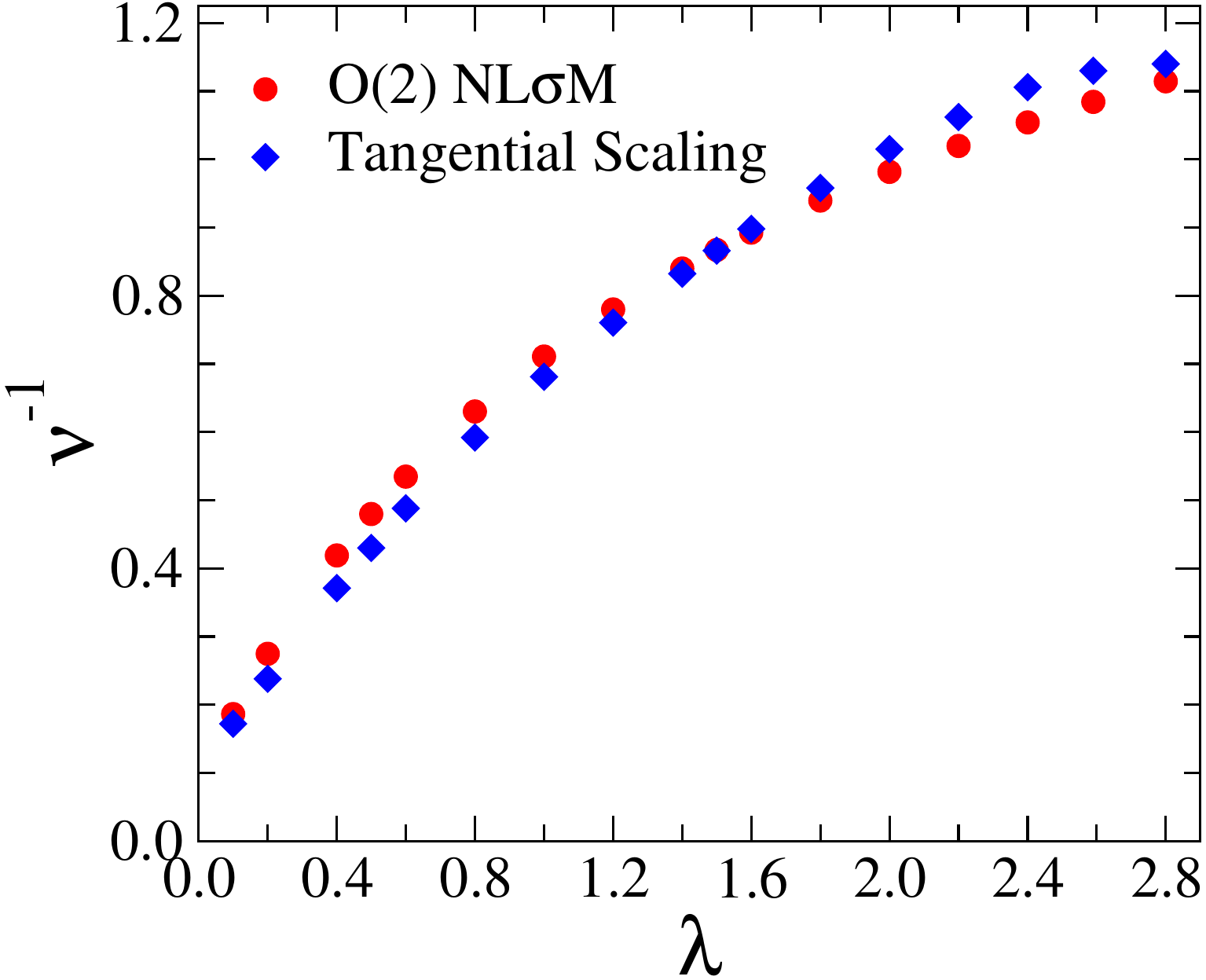}
    \caption{Inverse of the non-universal correlation length critical exponent along the Gaussian critical line. Data from the tangential finite-size scaling as well as the theoretical $O(2)$ NL$\sigma$M prediction are shown. Error bars are of the order of the symbol size.}
    \label{figE1}
\end{figure}

\begin{figure}[ht!]
    \centering
    \includegraphics[width=160pt, clip]{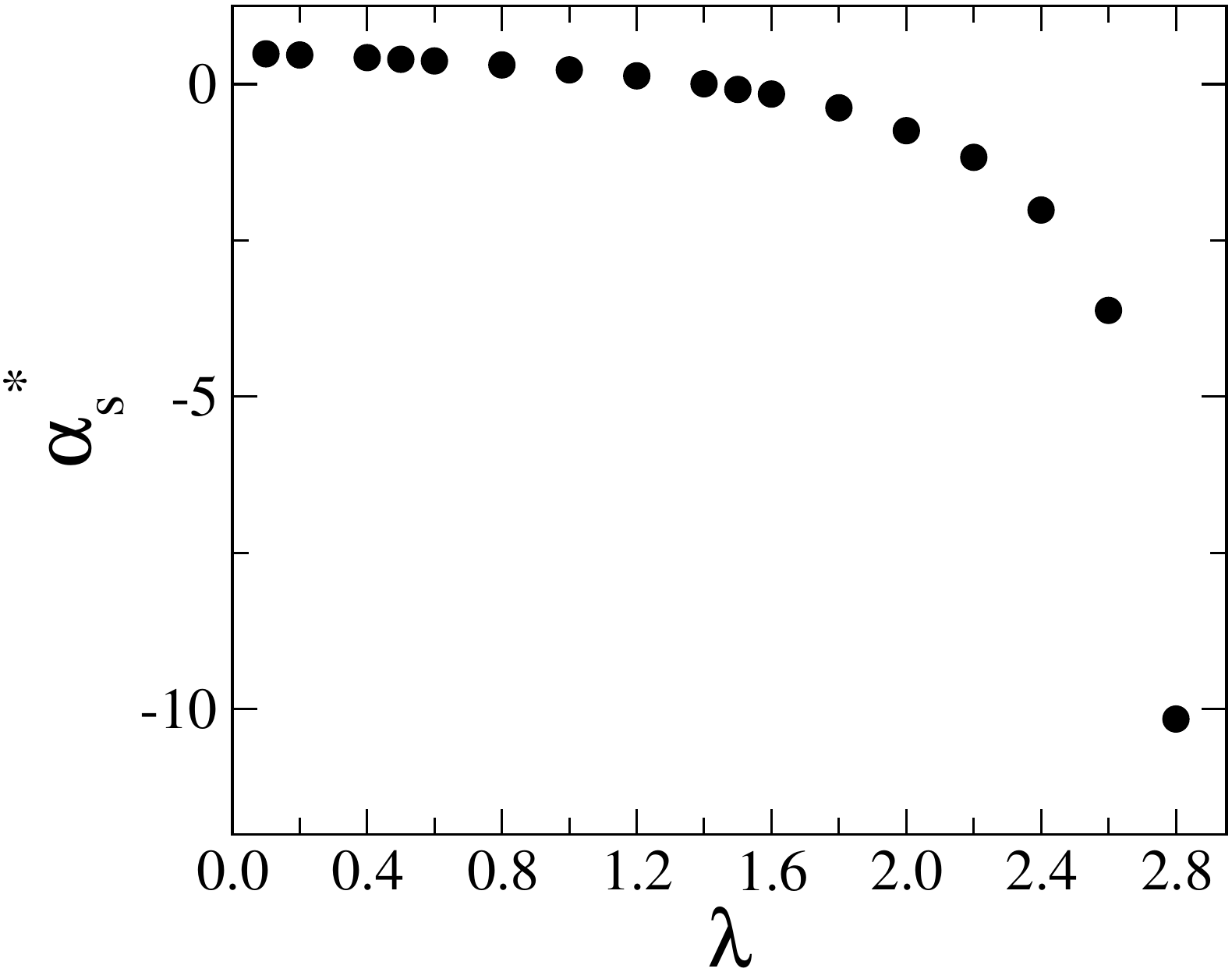}
    \caption{The first derivative of the scaled spin gap $\alpha_s^*=d\Delta_s/dD|_c$ along the Gaussian critical line. The values of $\alpha_s^*$ are positive for small $\lambda$, approaching to $\alpha_s^*=1/2$ as $\lambda \rightarrow 0$. Above $\lambda=1.4$, $\alpha_s^*$ becomes negative and diverges as the Gaussian critical line ends at a tricritical point \cite{Onc=1}.}
    \label{figE2}
\end{figure}

Despite the spin gap tangential scaling functions slightly differ along the Haldane to Large-D transition line (see Fig.\ref{figC}(d) and Fig.\ref{figD}(d)), the data collapse is successful  and reveals the Gaussian character of this critical line. The protocol put forward in the previous section was implemented along a wide range of $\lambda$ values. 
Besides the location of the critical line reported in Fig.\ref{figA}, we also report the estimated correlation length exponents in Fig.\ref{figE1}, as well as the scale invariant derivative of the scaled spin gap ${\alpha}_s^* = d\Delta_s/dD|_c$ along the critical line  (see  Fig.\ref{figE2}). All results are gathered in Table 1.

The estimated critical parameters $D_c$ and $\nu$ can be compared with available values found in the literature. In the Gaussian free quantum field model (a NL$\sigma$ M), the compactification radius parameter $K$ values are important in the determination of universality class of a quantum phase transition. It is related to the effective central charge $c=1$ of the Gaussian theory and the critical exponent $\nu = \frac{1}{(2-K)}$\cite{Onc=1}. Since $K=\pi/g$ and $g$ is given by Eq.\ref{eqg}, the Gaussian quantum field theory predicts that the correlation length exponent can be obtained if the location of the critical point is known, i.e., that these two critical quantities are not independent. Considering that the tangential finite-size scaling analysis provides independent measures of $D_c$ and $\nu$, we tested the NL$\sigma$M prediction by computing the values of $\nu$ from the expected values of the compactification radius $K$. The values of $1/\nu_{NL\sigma M}$ predicted by the NL$\sigma$M are also reported in Fig.\ref{figE1} and listed in the last column of Table 1. The agreement is fairly good, supporting the mapping on the NL$\sigma$M. The slight deviations at the ends of the critical line may be either due to the possible presence of relevant corrections to scaling due to the proximity of other critical points or an indicative that higher order terms in the field theoretical functional are needed to fully capture the critical behavior at the Haldane to Large-D transition.  As one decreases $\lambda$, the inverse of the correlation length exponent $1/\nu \rightarrow 0$, which indicates the Kosterlitz-Thouless nature of the transition at $\lambda=0$ with the compactification ratio $K = 2.0$\cite{PD,Onc=1}.

The derivative of the scaled spin gap along the Gaussian critical line shows some features that deserve attention, as shown in Fig.\ref{figE2}. For small  $\lambda$,  $\alpha_s^*$ has small but positive values, approaching to $\alpha_s^*=1/2$ for $\lambda\rightarrow 0$. Above $\lambda\simeq 1.4$, the critical scaled spin gap derivative changes sign and starts to diverge. It is interesting to recall that the Gaussian line meets the Ising critical line at a tricritical point.  The precise location of the tricritical point is still not completely settled\cite{PD,Onc=1}, although estimated to be close to
$D= 3.20$ and $\lambda = 2.90$. The tricritical point is described by the superposition of two conformal field theories with $c = \frac{1}{2}$ and $c = 1.0$ with the parameter $K$ approaching to $0.5$ from above\cite{Onc=1}.
The divergence of the scaled spin gap derivative as one approaches the tricritical point can be explored to obtain its precise location. However, due to the proximity of Gaussian and Ising critical points, larger system sizes than those considered in the present work would be required to properly deal with possible corrections to scaling. This specific topic will be left for a future larger scale computational effort.

In order to assess the accuracy of the present estimates of the Gaussian critical parameters based on the tangential finite-size scaling hypothesis, we compare some of our results with available high precision estimates.  A variant of the density matrix renormalization group method was used to study the Gaussian critical point at $\lambda =1.0$ and $\lambda= 0.5$ for the spin-1 model described by \ref{eq1}, reaching system sizes up to $N = 10000$ spins\cite{Gaussian1}. Exploring the peak behavior of von Neumann entropy, the  critical point for $\lambda=1.0$ was estimated to be $D_c = 0.96845(8)$. The approach of the gap to zero at $D_c$ from both Haldane and Large-D sides provided $\nu = 1.472(4)$. Both estimates are fully compatible with those found using the present tangential finite-size scaling, namely $D_c=0.9685(2)$ and $\nu=1.468(2)$. It is interesting to notice that the field theoretic relation  $\nu_{NL\sigma M}=1/(2-K)$ with $K=\pi/\sqrt{2(1+D+\lambda)}$ predicts the correlation length exponent $\nu_{NL\sigma M}=1.407$ for $(\lambda=1,D_c=0.9685)$. For $\lambda=0.5$, the critical point was estimated in Ref.\cite{Gaussian1} to be $D_c=0.6355(6)$ also in excellent agreement with the present estimate of $D_c=0.6352(1)$. In this case, the previous estimate of the correlation length exponent $\nu=2.387(5)$ was not derived using the scaling analysis of the closing gap but indirectly through the estimate of the compactification ratio $K$. This value is slightly above our present estimate of $\nu=2.325(22)$ based on the tangential scaling. Both estimates are above the field theoretic prediction $\nu_{NL\sigma M}=2.08$ for $(\lambda=0.5, D_c=0.6352)$. Our results are also compatible with the critical point ($\lambda=2.59 , D_c=2.30$) at which $1/\nu=1.15$ reported in Ref.\cite{Onc=1}.    

\begin{table}[h!]
    \centering 
    \begin{tabular}{|c|c|c|c|c|}
    \hline 
    $\lambda$ & $D_c$ & $\alpha_s^*$ & $\nu^{-1}$ & $\nu^{-1}_{NL \sigma M}$  \\ [0.5ex]
    \hline\hline
    0.10 & 0.4025(4) & 0.4831(1) & 0.172(3) & 0.1877(3)  \\
    0.20 & 0.4575(3) & 0.4648(4) & 0.238(6) & 0.2745(1) \\ 
    0.40 & 0.5741(2) & 0.4221(2) & 0.371(4) & 0.4189(1)  \\
    0.50 & 0.6352(2) & 0.3967(6) & 0.430(4) & 0.4797(1)   \\ 
    0.60 & 0.6982(3) & 0.3691(2) & 0.488(2) & 0.5346(1)   \\ 
    0.80 & 0.8298(4) & 0.3055(7) & 0.592(1) & 0.6301(1)   \\ 
    1.00 & 0.9685(2) & 0.2262(2) & 0.681(1) & 0.7107(1)   \\ 
    1.20 & 1.1139(3) & 0.1305(5) & 0.760(1) & 0.7797(1)   \\
    1.40 & 1.2658(1) & 0.0046(5) & 0.832(1) & 0.8397(1)   \\
    1.50 & 1.3440(2) & -0.092(8) & 0.866(1) & 0.8670(1)   \\
    1.60 & 1.4239(2) & -0.161(6) & 0.898(1) & 0.8926(1)   \\
    1.80 & 1.5881(1) & -0.381(8) & 0.958(1) & 0.9395(1)   \\
    2.00 & 1.7582(2) & -0.745(9) & 1.015(5) & 0.9816(1)   \\
    2.20 & 1.9343(1) & -1.171(9) & 1.062(3) & 1.0196(1)    \\
    2.40 & 2.1162(1) & -2.015(6) & 1.106(3) & 1.0542(1)   \\
    2.59 & 2.2944(2) & -3.62(8)  & 1.130(5)  & 1.0850(1)  \\
    2.80 & 2.4976(3) & -10.5(9)  & 1.14(1)  & 1.1148(1)   \\ [1ex]
    \hline
\end{tabular}
    
    \caption{Estimated values of the critical anisotropy $D_c$, first derivative of scaled gap at the critical point $\alpha_s^*$, and inverse of the correlation length critical exponent $\nu^{-1}$ obtained using the tangential finite-size scaling hypothesis for distinct values of $\lambda$ along the Haldane to Large-D transition. We also include the field-theoretic prediction of the correlation length critical exponent $\nu^{-1}_{NL\sigma M}$ based on the mapping on a NL$\sigma$M  given by $\nu^{-1} = 2 - \pi /\sqrt{2(1+D_c+\lambda)}$. For this, we used the estimated pairs $(\lambda, D_c)$.  }
\end{table}

\section{Concluding remarks}

In summary, we revisited the spin-1 XXZ Heisenberg model with uniaxial single-ion anisotropy aiming to characterize the Gaussian critical line using a specially tailored tangential finite-size scaling hypothesis. 
To clearly expose the distinct features of standard and tangential finite-size scaling, we focused in two transition lines. 

The first one was the Ising-like transition line between the gapless AF-Néel and the gaped Haldane phases for which standard finite-size scaling of the scaled neutral energy gap holds. The usual scaling hypothesis was shown to provide precise estimates of the critical line and the universal correlation length critical exponent $\nu=1$. 

The second transition investigated in the present work was between the gaped Haldane and large-D phases. This is a topological Gaussian phase-transition with the relevant energy spin gap closing at the critical point. The scaled energy gap and its derivative at the critical point are both scale invariant. To account for this specific feature, a tangential finite-size scaling hypothesis was implemented to properly write the scaled gap as a single function of the ratio between the chain size $N$ and the correlation length $\xi\propto|D-D_c|^{-\nu}$. 

Exploring the tangential scaling, we were able to locate the Gaussian transition and to estimate the non-universal correlation length critical exponent along the Gaussian critical line. The estimated values of the critical exponents were shown to be in good agreement with a field-theoretic prediction based in a mapping to a nonlinear sigma model\cite{Onc=1}. The estimated $D_c$ and $\nu$ values are also compatible with available data from the twisted boundary conditions method\cite{Kitazawa1,Kitazawa2} and high-accuracy calculations based on large-scale DMRG studies using chains up to $N=10^4$ sites\cite{Gaussian1}. Considering that the tangential finite-size scaling analysis requires data from relatively small chains sizes, it appears as a powerful procedure that adds to recent efforts aiming to investigate Gaussian topological quantum phase-transitions in general spin chains\cite{sc1,sc2,sc3,sc4,sc5}.

\section{Acknowlegdments}

This work was supported by CAPES (Coordenação de Aperfeiçoamente de Pessoal de Nivel Superior), CNPq (Conselho Nacional de Desenvolvimento Científico e Tecnológico), and FAPEAL (Fundação de Apoio à Pesquisa do Estado de Alagoas).

\medskip

\bibliographystyle{unsrt}
\typeout{}
\bibliography{library}

\end{document}